\documentclass[preprint,5p,times]{elsarticle}

\usepackage{amssymb}

\usepackage{amsfonts}
\usepackage{amsmath}
\usepackage{amssymb}
\usepackage{amsbsy}
\usepackage{amsthm}
\usepackage{epsfig}
\usepackage{graphicx}
\usepackage{colordvi}
\usepackage{graphics}
\usepackage{color}

\usepackage{rotating}
\usepackage{booktabs}
\usepackage{subfig}

\numberwithin{equation}{section}

\newcommand{\bfe}{{\bf e}}%
\newcommand{\bff}{{\bf f}}%
\newcommand{\bfk}{{\bf k}}%
\newcommand{\bfn}{{\bf n}}%
\newcommand{\bfr}{{\bf r}}%
\newcommand{\bfu}{{\bf u}}%
\newcommand{\bfv}{{\bf v}}%
\newcommand{\bfx}{{\bf x}}%
\newcommand{\bfomega}{\boldsymbol{\omega}}%
\newfont{\tenbfit}{cmbx10}%

\newfont{\tenbbb}{msbm10}%
\newfont{\svnbbb}{msbm8}%
\newcommand{\half}{{\textstyle{\frac{1}{2}}}}

\newcommand{\Curl}{\hbox{\rm curl}\mskip2mu}

\newcommand{\lj}{\mbox{$[\kern-0.1478125em[$}}
\newcommand{\rj}{\mbox{$]\kern-0.1478125em]$}}
\newcommand{\la}{\mbox{$\langle\kern-0.2325em\langle$}}
\newcommand{\ra}{\mbox{$\rangle\kern-0.2325em\rangle$}}
\newcommand{\Blj}{\mbox{$\Big[\kern-0.275em\Big[$}}
\newcommand{\Brj}{\mbox{$\Big]\kern-0.275em\Big]$}}
\newcommand{\Bla}{\mbox{$\Big\langle\kern-0.425em\Big\langle$}}
\newcommand{\Bra}{\mbox{$\Big\rangle\kern-0.425em\Big\rangle$}}

\newcommand{\zed}{{\bf 0}}

\newcommand{\captionfonts}{\footnotesize}

\makeatletter  % Allow the use of @ in command names
\long\def\@makecaption#1#2{%
  \vskip\abovecaptionskip
  \sbox\@tempboxa{{\captionfonts #1: #2}}%
  \ifdim \wd\@tempboxa >\hsize
    {\captionfonts #1: #2\par}
  \else
    \hbox to\hsize{\hfil\box\@tempboxa\hfil}%
  \fi
  \vskip\belowcaptionskip}
\makeatother   % Cancel the effect of \makeatletter

\begin{document}

\begin{frontmatter}



\title{Particle-based simulations of self-motile suspensions}


\author[label1]{Denis F.\ Hinz}
\ead{denis.hinz@mail.mcgill.ca}
\author[label2]{Alexander Panchenko}
\ead{panchenko@math.wsu.edu}
\author[label3]{Tae-Yeon Kim}
\ead{tykimsay@gmail.com}
\author[label3]{Eliot Fried\corref{cor1}}
\ead{mechanicist@gmail.com}

\cortext[cor1]{Corresponding author}
\address[label1]{Department of Mechanical Engineering, McGill University, Montr\'eal, Qu\'ebec, Canada H3A 2K6}
\address[label2]{Department of Mathematics, Washington State University, Pullman, WA 99164, USA}
\address[label3]{Department of Mechanical Engineering, University of Washington,  Seattle, WA 98195, USA}
%
%\fntext[fn1]{Department of Mechanical Engineering, McGill University, Montr\'eal, Qu\'ebec, Canada H3A 2K6}

%\address{}

\begin{abstract}
A simple model for simulating flows of active suspensions is investigated. The approach is based on dissipative particle dynamics. While the model is potentially applicable to a wide range of self-propelled particle systems, the specific class of self-motile bacterial suspensions is considered as a modeling scenario. To mimic the rod-like geometry of a bacterium, two dissipative particle dynamics particles are connected by a stiff harmonic spring to form an aggregate dissipative particle dynamics molecule. Bacterial motility is modeled through a constant self-propulsion force applied along the axis of each such aggregate molecule. The model accounts for hydrodynamic interactions between self-propelled agents through the pairwise dissipative interactions conventional to dissipative particle dynamics. Numerical simulations are performed using a customized version of the open-source LAMMPS (Large-scale Atomic/Molecular Massively Parallel Simulator) software package. Detailed studies of the influence of agent concentration, pairwise dissipative interactions, and Stokes friction on the statistics of the system are provided. The simulations are used to explore the influence of hydrodynamic interactions in active suspensions. For high agent concentrations in combination with dominating pairwise dissipative forces, strongly correlated motion patterns and a fluid-like spectral distributions of kinetic energy are found. In contrast, systems dominated by Stokes friction exhibit weaker spatial correlations of the velocity field. These results indicate that hydrodynamic interactions may play an important role in the formation of spatially extended structures in active suspensions.

\end{abstract}


\end{frontmatter}

% \linenumbers

%\section{}
%\label{}



%% References
\sloppy

%%%%%%%%%%%%%%%%%%%%%%%%%%%%%%%%%%%%%%%%%%%%%%%%
\section{Background and introduction}
%%%%%%%%%%%%%%%%%%%%%%%%%%%%%%%%%%%%%%%%%%%%%%%%

In comparison to their passive counterparts, active suspensions exhibit a wide range of non-classical phenomena. 
Such systems share the feature of self-motility which induces motions and mechanical stresses through the conversion of stored chemical-energy or ambient potential-energy into mechanical work. General descriptions and discussions of such systems can be found in the review articles by Pedley and Kessler~\cite{Pedley1992}, Ramaswamy~\cite{Ramaswamy2010}, Koch and Subramanian~\cite{Koch2011a}, and Marchetti et al.~\cite{Marchetti2013}.

One of the most ubiquitous phenomena observed in active suspensions is the presence of correlated large-scale motions over length scales which exceed those associated with the energy injected by active agents. Such patterns have been observed in experiments involving sessile droplets of self-motile bacterial suspensions (Dombrowski et al.~\cite{Dombrowski2004}, Tuval et al.~\cite{Tuval2005}, Cisneros et al.~\cite{Cisneros2007a}), two-dimensional free-standing self-motile bacterial films (Wu and Libchaber~\cite{Wu2001, Gregoire2001a, Wu2001a}, Sokolov et al.~\cite{Sokolov2007}, Sokolov and Aranson~\cite{Sokolov2009a, Sokolov2012}, Sokolov et al.~\cite{Sokolov2010}, and Liu and I~\cite{Liu2012}), quasi-two-dimensional thin layers of self-motile bacterial suspensions on surfaces (Zhang et al.\cite{Zhang2009a, Zhang2010}, Cisneros et al.~\cite{Cisneros2011}, Wensink et al.~\cite{Wensink2012c}, and Peruani et al.~\cite{Peruani2012}), as well as in bacterial suspensions in three-dimensional microfluidic chambers (Dunkel et al.~\cite{Dunkel2013} see also the associated viewpoint by Aranson~\cite{Aranson2013}). Moreover, spatially extended motion patterns similar to those observed in self-motile bacterial suspensions have been observed in motility assays consisting of protein filaments propelled by molecular motors (see, for example, N\'ed\'elec et al.~\cite{Nedelec1997}, Surrey et al.~\cite{Surrey2001}, Schaller et al.~\cite{Schaller2010}, Simuno et al.~\cite{Sumino2012}, Sanchez et al.~\cite{Sanchez2012}) and in the context of the swarming, herding, and flocking of fish, bird, or other animal colonies, as studied theoretically and numerically in the pioneering work of~Vicsek et al.~\cite{Vicsek1995}.

The observation of similar phenomena in active-matter systems for which disparate mechanisms underlie self-motility points at the presence of universal phenomena. However, the exact nature of the underlying kinematical and kinetic mechanisms leading to such phenomena are only partially understood. Further, to what extent these mechanisms share universal features seems unclear. 

Various different models aimed at reproducing, explaining, and predicting spatially extended motion in active suspensions have been proposed and investigated over the last decade. The present article focuses on agent-based models. Some agent-based approaches treat self-propelled agents as point particles that interact via with repulsive and attractive forces (see, for example, D'Orsogna et al.~\cite{DOrsogna2006}, Chuang et al.~\cite{Chuang2007}, and Carrillo et al.~\cite{Carrillo2009}). Other agent-based models rely on local and non-local interaction and alignment rules motivated from phenomenological observations in swarms and flocks (see, for example, Vicsek et al.~\cite{Vicsek1995}, Cucker and Smale~\cite{Cucker2007, Cucker2007a}, Carrillo et al.~\cite{Carrillo2010}, and Degond and Motsch~\cite{Degond2008}). Since the velocity and direction of motion of each individual self-propelled agent adjusts to the velocity and direction of its neighbours, numerical simulations based on such models exhibit behaviour in agreement with observed features of spatially extended correlated motion, flocking, or aggregation and alignment. However, the aggregation and alignment phenomena are {\emph{prescribed}} through phenomenological interaction rules. It is therefore unreasonable to expect such models to provide insights regarding the mechanisms {\emph{underlying}} these phenomena. 

A complimentary approach to modeling and understanding active suspensions is provided by agent-based models that do {\emph{not}} include prescribed alignment rules or attractive forces but rather rely on mechanically motivated forces, such as volume exclusion forces penalizing particle overlap, frictional forces accounting for interaction with the surrounding solvent, and random forces accounting for thermal fluctuations and intrinsic noise due to the swimming mechanisms of individual agents. These interaction forces are usually combined with the assumption that the self-propelled agents have rod-like geometry. 

For example, Peruani et al.~\cite{Peruani2006} found particle clustering for sufficiently large aspect ratios and rod concentrations in a two-dimensional model of self-propelled rods that interact solely through a soft volume exclusion force that penalizes overlap. Similar results were obtained with self-propelled rod models by Yang et al.~\cite{Yang2010}, who used a collection of shifted and truncated Lennard--Jones point potentials aligned along the axis of each rod, and by Wensink et al.~\cite{Wensink2012c} and Wensink and L\"owen~\cite{Wensink2012d}, who used a discrete set of repulsive Yukawa point potentials aligned along the axis of each rod. In these models, the agents are assumed to be immersed in an overdamped solvent modeled through Stokes-type friction forces that act on each agent. However, none of these models accounts for pairwise hydrodynamic interactions between self-propelled agents. A recent experimental investigation by Drescher et al.~\cite{Drescher2011} suggests that mechanical interactions between self-motile bacteria are dominated by short-range lubrication forces.

\graphicspath{ {./graphics//} }
 %%%%%%%%%
\begin{figure}[!t]
\begin{center}
%\begin{picture}(500,120)
\includegraphics[width=0.8\linewidth] {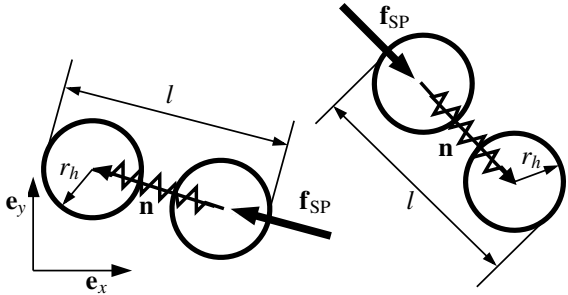}
\put(-60 ,30){$l$}
\put(-150,70){$l$}
\put(-100,30){$\bff_{\text{SP}}$}
\put(-70,100){$\bff_{\text{SP}}$}
\put(-17 ,47){$r_h$}
\put(-190 ,43){$r_h$}
\put(-180 ,0){$\bfe_x$}
\put(-210 ,30){$\bfe_y$}
\put(-160 ,28){$\bfn$}
\put(-48 ,50){$\bfn$}
%
%\end{picture}
\end{center}
\caption{Schematic of a pair of proximate dumbbells in the self-propelled soft-core dumbbell model. The vectors $\bfe_x$ and $\bfe_y$ are the orthogonal unit-valued basis vectors in the two-dimensional periodic domain.}
\label{DPDschematic01}
\end{figure}
%%%%%%%%%

To overcome this drawback of agent-based models, we propose a model based on purely repulsive self-propelled soft-core dumbbells with random forces and an additional type of interaction forces, namely pairwise dissipative forces. These forces are conventional ingredients of the dissipative particle dynamics (DPD) method, which is a mesoscopic particle-based simulation method widely used to study complex fluids and colloidal suspensions. Importantly, in agent-based models, where there is no explicit accounting for the suspending solvent, such pairwise dissipative interactions provide a simple means to account for hydrodynamic interactions between the agents. The inclusion of such pairwise dissipative forces in the DPD dumbbell model yields a novel model system to study the influence of hydrodynamic interaction forces in active suspensions.

The DPD method was pioneered by Hoogerbrugge and Koelman~\cite{Hoogerbrugge1992, Koelman1993}. As a mesoscale particle simulation method, it was originally designed to explore hydrodynamic behavior in complex fluids, as discussed, for example, in a review article by Moeendarbary et al.~\cite{Moeendarbary2009}. Since their governing equations are usually simpler than the partial-differential equations arising from suitable continuum alternatives, mesoscale particle simulation methods are promising tools for simulating complex fluid flows in nonsimple geometries. Further, structure of the equations of the DPD method resembles that of molecular dynamics (MD) and, thus, can be solved efficiently in existing MD codes. Moreover, as Groot and Warren~\cite{Groot1997} explain, the use of soft-core interaction potentials in the DPD method allows for large integration timesteps and stable simulations, providing a significant speedup in comparison to MD simulations with hard-core interaction potentials.

While the DPD method has a number of other desirable properties, such as, for example, net-momentum conservation and Galilean invariance (see, for example, Allen and Schmid \cite{Allen2007}), the ability to produce hydrodynamic behaviour with computational efficiency also makes it a potentially powerful tool for the simulation of active suspensions. An active Brownian point-particle model with supplemental dissipative DPD-type interactions was recently studied by Lobaskin and Romenskyy~\cite{Lobaskin2013}. They report a transition to an ordered ``swarming" state with increasing particle concentration and energy input through increasing magnitude of the self-propulsion force. Based on comparison of correlation functions and order parameters obtained from their model with these of the Vicsek model~\cite{Vicsek1995}, Lobaskin and Romenskyy~\cite{Lobaskin2013} argue that pairwise dissipative interactions can be seen as an alternative alignment mechanism. 

In the present article, the DPD method is adopted to study a system of self-propelled agents. While the model is potentially applicable to a wide range of self-propelled particle systems, the specific class of self-motile bacterial suspensions is considered as a modeling scenario. Using this model, the central aim of this study is to investigate the influence of hydrodynamic interaction forces. Particular attention is placed on the influence of Stokes-type friction forces and pairwise dissipative interactions on the behavior of the system along with limiting cases wherein either of the two dissipative forces dominates. Such a generic study of hydrodynamic interaction forces is meant to assess the importance of different interaction mechanisms.

This article is structured as follows. In Section~\ref{sec:SPPModel}, the self-propelled soft-core dumbbell model is introduced. 
In Section~\ref{sec:Nondim}, the salient dimensionless parameters of the model are derived. In Section~\ref{sec:forces}, the roles of the pairwise interaction forces and non-pairwise forces are explained in detail. In Section~\ref{sec:SPPnumericalStudy}, numerical simulations are performed to examine the influences of pairwise and non-pairwise dissipative forces on the statistics of the system in a two-dimensional square domain with periodic boundary conditions. Finally, a summary and concluding remarks are provided in Section~\ref{sec:SPPsummary}.

%%%%%%%%%%%%%%%%%%%%%%%%%%%%%%%%%%%%%%%%%%%%%%%%
\section{Self-propelled soft-core dumbbell model}
\label{sec:SPPModel}
%%%%%%%%%%%%%%%%%%%%%%%%%%%%%%%%%%%%%%%%%%%%%%%%

The motion of a particle $i$ with constant particle mass $m^{i}$ in a system of $N$ particles is governed by Newton's equations 
\begin{equation}\label{eq:Newton}
\left.
\begin{array}{l}
\dot \bfx^{i} = \bfv^{i},\\
m^{i}\dot \bfv^{i} = \sum\limits_{j=1}^N\bff^{ij}+ \bff^{i}_{e},
\end{array}
\!\!
\right\}
\end{equation}
where a superposed dot denotes time differentiation, $\bfx^{i}$ and $\bfv^{i}$ denote the location and velocity of particle $i$, $\bff^{ij} = - \bff^{ji}$ is the pairwise particle interaction force between particles $i$ and $j$, and $\bff^{i}_{e}$ is the external force excerted on particle $i$.

%%%%%%%%%%%%%%%%%%%%%%%%%%%%%%%%%%%%%%%%%%%%%%%%
\subsection{Pairwise interaction forces}
%%%%%%%%%%%%%%%%%%%%%%%%%%%%%%%%%%%%%%%%%%%%%%%%

The pairwise interaction forces $\bff^{ij}$ of the model are based on the pairwise interaction forces of the DPD method (see, for example, Groot and Warren~\cite{Groot1997})
\begin{equation}\label{eq:DPDf}
%\begin{array}{ll}
\bff^{ij} = \begin{cases} \bff^{ij}_{\rm{C}}+\bff^{ij}_{\rm{D}}+\bff^{ij}_{\rm{R}},  & \quad r<r_c, \\
 \zed,   & \quad \text{otherwise},
\end{cases}
%\end{array}
\end{equation}
where $\bff^{ij}_{\text{C}}$ is purely conservative force, $\bff^{ij}_{\text{D}}$ is a purely dissipative force, $\bff^{ij}_{\text{R}}$ is a random force, $r=|\bfr^{i}-\bfr^{j}|$ is the distance between particles $i$ and $j$, and $r_c$ is the cutoff radius for pairwise interactions. Following Groot and Warren \cite{Groot1997}, these forces are given as 
\begin{equation}\label{eq:DPDcdr}
\left.
\begin{array}{l}
\bff^{ij}_{\text{C}}= Aw(r) \hat{\bfe}^{ij},
\\
\bff^{ij}_{\text{D}}= -\gamma w^2(r) [(\bfv^{i}-\bfv^{j})\cdot \hat{\bfe}^{ij}]\hat{\bfe}^{ij},
\\
\bff^{ij}_{\text{R}}= \sqrt{2\gamma k_B T} w(r) \alpha^{ij} (\Delta t)^{-1/2} \hat{\bfe}^{ij},
\end{array}
\!\!
\right\}
\end{equation}
in which $w$ defined such that
\begin{equation}\label{eq:weightFunc01}
w(r) = 1-\frac{r}{r_c}
\end{equation}
 is the soft-core weighting function, $\hat{\bfe}^{ij} = (\bfr^{i}- \bfr^{j}) / r$ is the unit vector directed from particle $i$ to particle $j$, $\alpha^{ij}$ is a random number, $A > 0$ is the conservative force parameter, and $\gamma >0$ is the pairwise friction parameter. Based on the assumption that all particle interactions are governed by the same parameters, $A$ and $\gamma$ are constant and independent of $i$ and $j$. In~\eqref{eq:DPDcdr}$_3$, $k_B T$, with $k_B$ and $T$ Boltzmann's constant and the absolute temperature, provides a reference energy scale and $\Delta t$ is the simulation timestep. Together with the pairwise friction parameter $\gamma$, $k_B T$ characterizes the magnitude of the pairwise random forces accounting for Brownian fluctuations.

%%%%%%%%%%%%%%%%%%%%%%%%%%%%%%%%%%%%%%%%%%%%%%%%
\subsection{Dumbbell bond forces}
%%%%%%%%%%%%%%%%%%%%%%%%%%%%%%%%%%%%%%%%%%%%%%%%

To mimic the rod-like geometry of a bacterium, two DPD particles are connected by a stiff harmonic spring to form an aggregate DPD molecule, as depicted schematically in Figure~\ref{DPDschematic01}. The bond between two DPD particles forming a single dumbbell is modeled with a harmonic spring potential (see, for example, Schwarz--Linek et al.~\cite{Schwarz-Linek2012}) $V_H$ of the form 
\begin{equation}\label{eq:stiffSpring}
V_H(r) =\half \kappa(r-l)^2,
\end{equation}
where $\kappa>0$ is the constant spring stiffness and $l$ is the equilibrium length between two bonded particles. Notice that $\kappa$ must be chosen consistent with the relatively inextensible nature of a bacterium. Since two particles form a single agent, the bond forces are considered exclusively between these two particles and other possible interaction forces are switched off.

%%%%%%%%%%%%%%%%%%%%%%%%%%%%%%%%%%%%%%%%%%%%%%%%
\subsection{Non-pairwise forces}
%%%%%%%%%%%%%%%%%%%%%%%%%%%%%%%%%%%%%%%%%%%%%%%%

Apart from standard pairwise DPD forces~\eqref{eq:DPDcdr} and dumbbell bond forces associated with the harmonic spring potential~\eqref{eq:stiffSpring}, self-propulsion and Stokes friction are incorporated through non-pairwise forces 
\begin{equation}\label{eq:DPDnp}
%\left.
%\begin{array}{l}
\displaystyle\bff^{k}_{\rm{SP}}= f_{\rm{SP}} \bfn^{k} \qquad\text{and}\qquad
\displaystyle \bff_{\rm{S}}^{i} = - \gamma_s \bfv^{i},
%\end{array}
%\!\!
%\right\}
\end{equation}
with $\gamma_s \ge 0 $ the constant Stokes friction parameter and $f_{\rm{SP}}\ge 0 $ the magnitude of the constant self-propulsion force along the unit-valued director $\bfn^{k}$ of each agent $k$, as shown in Figure~\ref{DPDschematic01}. The unit-valued director $\bfn^{k}$ is defined through the location vectors of the particles $i$ and $j$ forming the single agent $k$
\begin{equation}
\bfn^k = \frac{\bfr^i - \bfr^j}{|\bfr^i - \bfr^j|}.
\end{equation}
Notice that indices $i$ and $j$ are used to refer to particles, whereas the index $k$ is used to refer to one agent consisting of two particles. 

Importantly, the model involves two types of dissipative forces: 
\begin{enumerate}
\item The Stokes friction $\bff_{\rm{S}}^{i}$ acting on particle $i$.
\item The dissipative interaction force $\bff_{\rm{D}}^{ij}$ defined in~\eqref{eq:DPDcdr}$_2$ acting between a pair of particles $i$ and~$j$.
\end{enumerate}
While the former is proportional to the velocity $\bfv^{i}$ of particle $i$, the latter is proportional to the relative velocity $\bfv^i-\bfv^j$ of the two particles and the soft-core weighting function $w$ defined in~\eqref{eq:weightFunc01}. The magnitudes of the dissipative forces are determined by the friction parameters $\gamma$ and $\gamma_s$, which correspond respectively to the pairwise dissipative interaction force and the Stokes friction.

%%%%%%%%%%%%%%%%%%%%%%%%%%%%%%%%%%%%%%%%%%%%%%%%
\section{Nondimensionalization}
\label{sec:Nondim}
%%%%%%%%%%%%%%%%%%%%%%%%%%%%%%%%%%%%%%%%%%%%%%%%

To obtain a dimensionless version of the formulation, introduce the characteristic length $l_0$ of a bacterium (as modeled by a dumbbell), a characteristic magnitude $f_0$ of the self-propulsion force, and a characteristic friction parameter $\gamma_0$. 
Define dimensionless counterparts $x^*$, $v^*$, $t^*$, $m^*$, and $f^*$ of $x$, $v$, $t$, $m$, and $f$ by
\begin{equation}\label{char1}
\begin{array}{l}
\displaystyle x^* = \frac{x}{l_0}, \qquad v^* = \frac{v \gamma_0 }{f_0}, \qquad t^* = \frac{t f_0}{\gamma_0 l_0},  \\
\displaystyle m^* = \frac{m f_0}{\gamma_0^2 l_0},\qquad\text{and}\qquad f^* = \frac{f}{f_0}.
\end{array}
\end{equation}
Since the equations~\eqref{eq:DPDcdr} involve the product $k_B T$, which carries dimensions of energy, there is no need to explicitly include a temperature in the nondimensionalization.

Applying~\eqref{char1} to~\eqref{eq:Newton} yields the dimensionless equations of motion,
\begin{equation}\label{eq:NewtonND}
\left.
\begin{array}{l}
\dot \bfx^{*i} = \bfv^{*i},\\
m^{*i}\dot \bfv^{*i} = \sum\limits_{j=1}^N\bff^{*ij}+ \bff^{*i}_{e},
\end{array}
\!\!
\right\}
\end{equation}
where a superposed dot now denotes differentiation with respect to the nondimensional time $t^*$. Similarly, in view of~\eqref{eq:DPDcdr} and~\eqref{char1}, the dimensionless versions of the pairwise forces are given by 
\begin{equation}\label{eq:DPDcdrND}
\left.
\begin{array}{l}
\bff^{*ij}_{\text{C}}= \frac{A}{f_0}w(r^*) \hat{\bfe}^{ij},
\\
 \bff^{*ij}_{\text{D}}= -\frac{\gamma}{\gamma_0} w^2(r^*) [(\bfv^{*i}-\bfv^{*j})\cdot \hat{\bfe}^{ij}]\hat{\bfe}^{ij},
\\
\bff^{*ij}_{\text{R}}= \sqrt{2\frac{\gamma}{\gamma_0} \frac{k_B T}{f_0 l_0}} w(r^*) \alpha^{ij} (\Delta t^*)^{-1/2} \hat{\bfe}^{ij},
\end{array}
\!\!
\right\}
\end{equation}
and the dimensionless self-propulsion force and Stokes friction are given by
\begin{equation}\label{eq:DPDnpND}
%\left.
%\begin{array}{l}
\bff^{*k}_{\text{SP}}= \frac{f_{\text{SP}}}{f_0} \bfn^{k}\qquad\text{and}\qquad
 \bff_{\text{S}}^{*i} = -\frac{ \gamma_s}{\gamma_0} \bfv^{*i}.
%\end{array}
%\!\!
%\right\}
\end{equation}
Inspection of~\eqref{eq:DPDcdrND} and~\eqref{eq:DPDnpND} identifies the following important dimensionless parameters:
\begin{equation}\label{eq:NDparameters}
\begin{array}{l}
\displaystyle \Phi = \frac{A}{f_0} >0, \qquad {\rm{Pe}} = \frac{f_0 l_0}{k_B T} > 0, \\ 
\displaystyle \Gamma = \frac{\gamma}{\gamma_0} \ge 0,  \qquad \Gamma_s = \frac{ \gamma_s}{\gamma_0} \ge 0.
\end{array}
\end{equation}

While $\Phi$ characterizes the ratio of conservative pairwise interaction forces to self-propulsion forces, $\text{Pe}>0$ is an ``active" P\'eclet number (see Schwarz-Linek et al.~\cite{Schwarz-Linek2012}) characterizing the importance of self-propulsion energy relative to the energy associated with random forces due to thermal fluctuations and intrinsic fluctuations of the swimming mechanisms of self-propelled agents. For $\text{Pe}\ll 1$, random forces dominate over self-propulsion forces. The limit $\rm{Pe}\rightarrow \infty$ encompasses the scenario in which random forces are negligible in comparison to self-propulsion forces. The dimensionless friction parameters $\Gamma$ and $\Gamma_s$ respectively characterize pairwise dissipative interactions and Stokes-type dissipative forces. 

With~\eqref{eq:NDparameters}, the model explicitly allows for the limiting cases of vanishing Stokes friction or vanishing pairwise dissipative interactions. More particularly, the limiting case $\Gamma \rightarrow 0$ represents the limit of negligible hydrodynamic interaction forces and dominant Stokes friction. Conversely, the limit case of $\Gamma_s \rightarrow 0$ represents the limit case of negligible Stokes friction and dominant pairwise hydrodynamic interaction forces. From a physical perspective, it is reasonable to expect that while Stokes friction should dominate in the regime of dilute agent concentrations, the influence of pairwise interaction forces become significant with increasing agent concentration.

On substituting~\eqref{eq:NDparameters} into~\eqref{eq:DPDcdrND} and~\eqref{eq:DPDnpND}, the pairwise and non-pairwise forces become  
\begin{equation}\label{eq:DPDcdrND2}
\left.
\begin{array}{l}
\bff^{*ij}_{\text{C}}= \Phi w(r^*) \hat{\bfe}^{ij},
\\
 \bff^{*ij}_{\text{D}}= -\Gamma w^2(r^*) [(\bfv^{*i}-\bfv^{*j})\cdot \hat{\bfe}^{ij}]\hat{\bfe}^{ij},
\\
\bff^{*ij}_{\text{R}}= \sqrt{2\Gamma \text{Pe}^{-1} } w(r^*) \alpha^{ij} (\Delta t^*)^{-1/2} \hat{\bfe}^{ij},
\end{array}
\!\!
\right\}
\end{equation}
and 
\begin{equation}\label{eq:DPDnpND2}
\left.
\begin{array}{l}
\bff^{*k}_{\text{SP}}= f^*_{\text{SP}} \bfn^{k},
\\
 \bff^{*i}_{\text{S}} = -\Gamma_s \bfv^{*i}, \\
\end{array}
\!\!
\right\}
\end{equation}
respectively. For convenience, asterisks appearing in~\eqref{eq:NewtonND},~\eqref{eq:DPDcdrND2}, and~\eqref{eq:DPDnpND2} are omitted hereafter.

%%%%%%%%%%%%%%%%%%%%%%%%%%%%%%%%%%%%%%
\section{The role of the different forces}
\label{sec:forces}
%%%%%%%%%%%%%%%%%%%%%%%%%%%%%%%%%%%%%%

The model involves a collection of pairwise and non-pairwise interaction forces~\eqref{eq:DPDcdrND2} and~\eqref{eq:DPDnpND2}. The nature of these forces  may be summarized as follows:
\begin{enumerate}
\item The self-propulsion force $\bff^{k}_{\text{SP}}$: This non-pairwise entity acts on each agent and accounts for the motility of each agent $k$ contained in the active suspension.
\item The Stokes friction $\bff_{\text{S}}^{i}$: This non-pairwise entity accounts for hydrodynamic forces exerted by the surrounding solvent, acting on each particle $i$ forming the aggregate agent $k$ in the active suspension.
\item The pairwise conservative force $\bff_{\text{C}}^{ij}$: This pairwise entity acts as a soft volume-exclusion force. Together with the cutoff radius, this force defines the geometry and size of each particle~$i$. (See also the discussion of the hydrodynamic radius of a DPD particle in Section~\ref{sec:param}.)
\item The pairwise random force $\bff_{\text{R}}^{ij}$: This pairwise entity accounts for thermal fluctuations due to Brownian motion and intrinsic fluctuations in the swimming mechanisms of each agent.
\item The pairwise dissipative interaction force $\bff_{\text{D}}^{ij}$: This pairwise entity accounts for dissipative hydrodynamic interaction forces distinct from those encompassed by non-pairwise Stokes friction. Since this force is proportional to velocity differences of interacting particle pairs, it formally resembles a lubrication force (see, for example, Kim and Karrila~\cite{Kim1991}). Further, the pairwise dissipative forces~\eqref{eq:DPDcdrND2}$_2$ of the model depend quadratically on the dimensionless separation distance between interacting particle pairs through the square $w^2$ of the soft-core weighting function $w$, which differs from the dependence of lubrication forces on a normalized separation distance. Lubrication forces of two approaching spheres possess a hard-core functional dependence on the normalized separation distance $\epsilon$ between the spheres that is proportional to $\epsilon^{-1}$ as $\epsilon\rightarrow 0$.  The model under consideration is intended to characterize the influence of such pairwise dissipative forces in a general context inspired by the DPD framework. At this stage, the model does not account for alternative kinds of weighting functions. However, the impact of choosing different weighting functions should be a subject of future research. 
\end{enumerate}

Being a purely mechanical theory based on pairwise interaction forces and non-pairwise forces, the present model deviates from other available agent-based models. 

Most importantly, the model does not include a priori ``biological alignment" rules or ad-hoc coordination forces. These ingredients are often incorporated through long-range attraction forces or specifically prescribed alignment or clustering requirements. Here, the only interaction forces are short-ranged repulsive, random,  and dissipative; in particular, no long-range attraction forces are introduced. Further, it is left to the underlying dynamics of the system to determine the extent to which alignment or clustering develops. 

The approach taken here allows for investigations of the statistical behaviour of a  generic self-propelled particle model system without biological rules, but with pairwise dissipative hydrodynamic interaction forces. The ensuing investigations are aimed at revealing universal features of self-propelled particle systems. While biological alignment and clustering rules or coordination forces might well exist, the characterization of such rules seems elusive, at least at present. The model used in the present work cannot be expected to capture phenomena associated with such rules or forces. However, it enables a clear separation of purely mechanical phenomena from other erstwhile biological phenomena and provides a basis for characterizing regimes in which certain phenomena might dominate over other phenomena.

%%%%%%%%%%%%%%%%%%%%%%%%%%%%%%%%%%%%%%
\section{Numerical study}
\label{sec:SPPnumericalStudy}
%%%%%%%%%%%%%%%%%%%%%%%%%%%%%%%%%%%%%%

In this section, results from numerical simulations based on the equations of motion~\eqref{eq:NewtonND} with the pairwise interaction forces~\eqref{eq:DPDcdrND2} and non-pairwise forces~\eqref{eq:DPDnpND2} along with the stiff harmonic spring constraint~\eqref{eq:stiffSpring} are presented. These simulations are conducted in a two-dimensional square domain with periodic boundary conditions. Simulations are performed using a customized version of the molecular dynamics package LAMMPS~\cite{Plimpton1995, LAMMPS2013}. The standard velocity-Verlet~\cite{Verlet1967} time integration scheme with a dimnesionless integration timestep of $\Delta t = 3.0\cdot 10^{-3}$ is used for all simulations. Initially, the agents are taken to be randomly distributed and randomly oriented throughout the computational domain. The initial velocities of the agents are set to zero.  
Focus is on the influences of the volume fraction $\phi$ and two friction parameters $\Gamma$ and $\Gamma_s$ on the statistics of the system. All results presented are from the (statistically) steady-state regime and represent space-time averages. To compute statistical objects, snapshots of the agent and velocity distribution at different timesteps in the steady-state regime are collected every $2.0\cdot 10^4$ timesteps. Next, statistical objects computed for at least 10 such snapshots are averaged, the resulting statistical object representing a space-time average.

%%%%%%%%%%%%%%%
\begin{figure*}[t!]
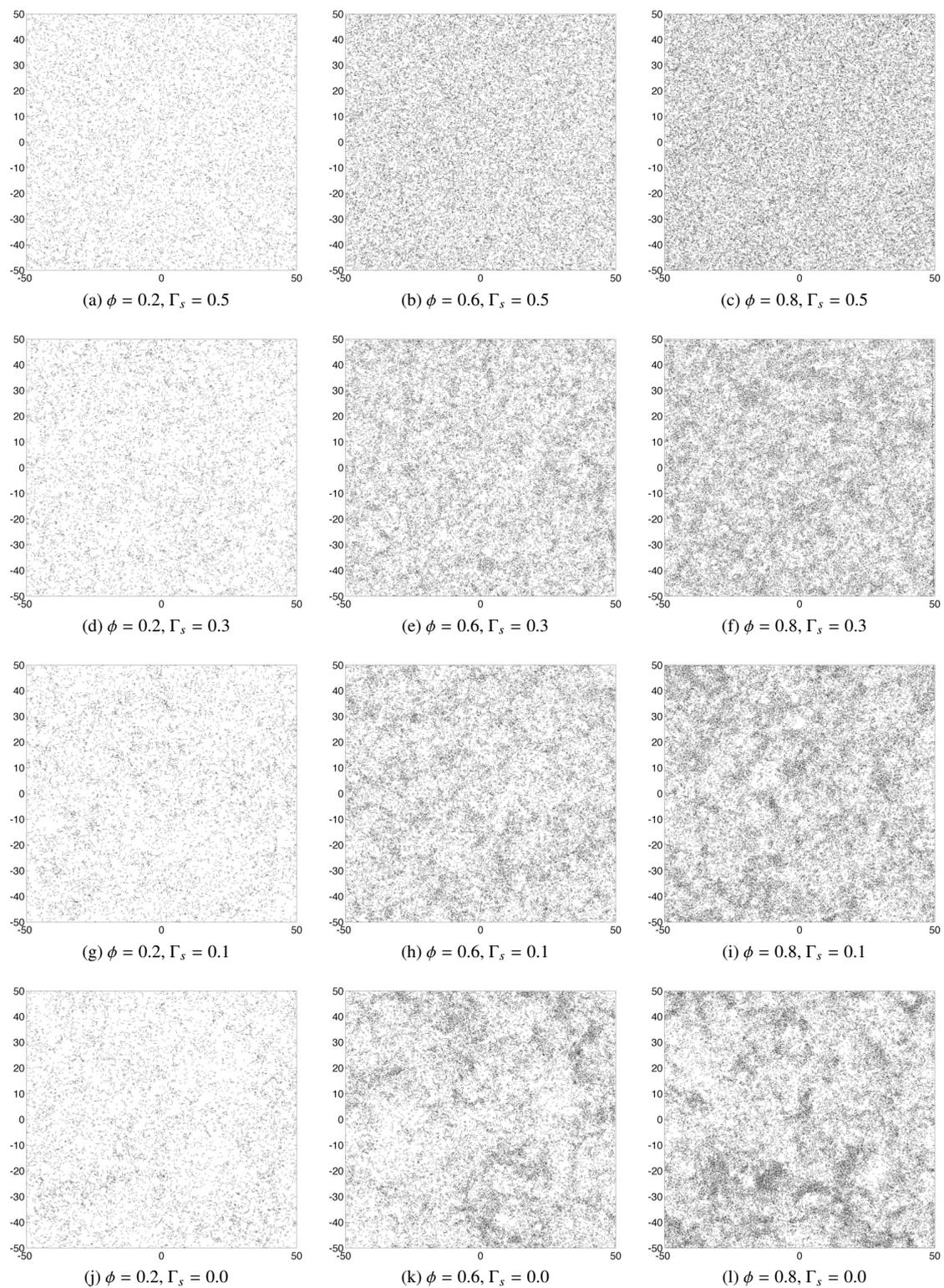

\begin{center}

%\begin{picture}(500,550)
\graphicspath{ {./graphics/serie07_rod_postprocessing/postResults/serie07_T_0p05_prop_1p0_r_0p5_a1p5_g0p5//} }
\subfloat[$\phi = 0.2$, $\Gamma_s = 0.5$\label{subfig-1:dummy}]{%
\includegraphics[width=.25\textwidth] {phi_0p2_l_0p75_Gamma_8p0/ROD_2000000.jpg}}
$\quad$
\subfloat[$\phi = 0.6$, $\Gamma_s = 0.5$\label{subfig-1:dummy}]{%
\includegraphics[width=.25\textwidth] {phi_0p6_l_0p75_Gamma_8p0/ROD_2000000.jpg}}
$\quad$
\subfloat[$\phi = 0.8$, $\Gamma_s = 0.5$\label{subfig-1:dummy}]{%
\includegraphics[width=.25\textwidth] {phi_0p8_l_0p75_Gamma_8p0/ROD_2000000.jpg}}
\graphicspath{ {./graphics/serie07_rod_postprocessing/postResults/serie07_T_0p05_prop_1p0_r_0p5_a1p5_g0p3//} }
\subfloat[$\phi = 0.2$, $\Gamma_s = 0.3$\label{subfig-1:dummy}]{%
\includegraphics[width=.25\textwidth] {phi_0p2_l_0p75_Gamma_8p0/ROD_2000000.jpg}}
$\quad$
\subfloat[$\phi = 0.6$, $\Gamma_s = 0.3$\label{subfig-1:dummy}]{%
\includegraphics[width=.25\textwidth] {phi_0p6_l_0p75_Gamma_8p0/ROD_2000000.jpg}}
$\quad$
\subfloat[$\phi = 0.8$, $\Gamma_s = 0.3$\label{subfig-1:dummy}]{%
\includegraphics[width=.25\textwidth] {phi_0p8_l_0p75_Gamma_8p0/ROD_2000000.jpg}}
\graphicspath{ {./graphics/serie07_rod_postprocessing/postResults/serie07_T_0p05_prop_1p0_r_0p5_a1p5_g0p1//} }
\subfloat[$\phi = 0.2$, $\Gamma_s = 0.1$\label{subfig-1:dummy}]{%
\includegraphics[width=.25\textwidth] {phi_0p2_l_0p75_Gamma_8p0/ROD_2000000.jpg}}
$\quad$
\subfloat[$\phi = 0.6$, $\Gamma_s = 0.1$\label{subfig-1:dummy}]{%
\includegraphics[width=.25\textwidth] {phi_0p6_l_0p75_Gamma_8p0/ROD_2000000.jpg}}
$\quad$
\subfloat[$\phi = 0.8$, $\Gamma_s = 0.1$\label{subfig-1:dummy}]{%
\includegraphics[width=.25\textwidth] {phi_0p8_l_0p75_Gamma_8p0/ROD_2000000.jpg}}
\graphicspath{ {./graphics/serie07_rod_postprocessing/postResults/serie07_T_0p05_prop_1p0_r_0p5_a1p5_g0p0//} }
\subfloat[$\phi = 0.2$, $\Gamma_s = 0.0$\label{subfig-1:dummy}]{%
\includegraphics[width=.25\textwidth] {phi_0p2_l_0p75_Gamma_8p0/ROD_2000000.jpg}}
$\quad$
\subfloat[$\phi = 0.6$, $\Gamma_s = 0.0$\label{subfig-1:dummy}]{%
\includegraphics[width=.25\textwidth] {phi_0p6_l_0p75_Gamma_8p0/ROD_2000000.jpg}}
$\quad$
\subfloat[$\phi = 0.8$, $\Gamma_s = 0.0$\label{subfig-1:dummy}]{%
\includegraphics[width=.25\textwidth] {phi_0p8_l_0p75_Gamma_8p0/ROD_2000000.jpg}}
\end{center}
\caption{Agent distribution fields in the dilute ($\phi = 0.2$), semidilute ($\phi = 0.6$), and dense ($\phi = 0.8$) regimes for $\Gamma = 8.0$ and different values of $\Gamma_s$, including the limiting case of $\Gamma_s=0.0$. }
\label{fig:particleField01}
\end{figure*}
%%%%%%%%%%%%%%%

%%%%%%%%%%%%%%%%%%%%%%%%%%%%%%%%%%%%%%%%%%%%%%%%
\subsection{Representative parameter values}
\label{sec:param}
%%%%%%%%%%%%%%%%%%%%%%%%%%%%%%%%%%%%%%%%%%%%%%%%

The parameters entering the equations of motion along with pairwise and non-pairwise forces are determined from estimates and references in the literature. 

Apart from the cutoff radius $r_c$, Pan et al.~\cite{Pan2008} observed that the characteristic linear dimension of an individual DPD particles may be distiguished through a nondimensional hydrodynamic radius $r_h<r_c$. The quantity $r_h$ and the nondimensional length $l$ of a single agent (Figure~\ref{DPDschematic01}) are used to define the aspect ratio $a = l/2r_h$; for illustrative purposes, it is assumed that $l=1.0$ and $a=4.0$. Following Pan et al.~\cite{Pan2008}, the nondimensional cutoff radius takes the value $r_c=0.5$. 

Using the choice $r_c = 0.5$, the magnitude of the conservative interaction parameter $\Phi$ may be estimated by adopting the view that, together with the nondimensional cutoff radius, the magnitude of the repulsive volume-exclusion force~\eqref{eq:DPDcdrND2}$_1$ defines of the linear dimensions of a single particle. In other words, the repulsive volume-exclusion force is expected to (statistically) balance other forces acting on the particle  to prevent particle overlap beyond the hydrodynamic radius $r_h$. Granted that the self-propulsion forces are dominant, an approximate balance between the self-propulsion force~\eqref{eq:DPDnpND2}$_1$ and the magnitude of the conservative volume-exclusion force~\eqref{eq:DPDcdrND2}$_1$ of two isolated interacting particles leads to the estimate $f_{\text{SP}} \approx  \Phi w(r=r_h)$ and, thus to
\begin{equation}\label{eq:fbalance02}
\Phi \approx  \frac{f_{\text{SP}}}{1-r_h/r_c},
\end{equation}
where inertia and other forces have been neglected; the estimate~\eqref{eq:fbalance02} gives an approximate relationship between the nondimensional self-propulsion force and the nondimensional repulsive volume-exclusion force depending on the nondimensional cutoff radius $r_c$ and the nondimensional hydrodynamic radius $r_h$. With $r_c$, $r_h$, and the choice $f_{\text{SP}}=1.0$, \eqref{eq:fbalance02} gives the estimate $\Phi \approx 1.3$. Since \eqref{eq:fbalance02} accounts for only two isolated particles and neglects inertia and all other interaction forces, it should be interpreted as a lower bound for $\Phi$; it hence seems reasonable to choose $\Phi = 1.5$.

Furthermore, the value of the active P\'eclet number is taken to be ${\rm{Pe}}=2.0\cdot 10^1$, as estimated by Schwarz-Linek et al.~\cite{Schwarz-Linek2012}. Since bacteria are nearly inextensible, a large value $\kappa=1.0 \cdot 10^3$ for the nondimensional spring stiffness is selected to ensure that the length of the agents remains practically constant.

With all the previously discussed parameters set, the remaining free parameters are the friction coefficients $\Gamma$ and $\Gamma_s$ along with the concentration of the bacteria. To measure the latter quantity, a nondimensional area fraction
\begin{equation}
\phi = \frac{N l 2r_h}{{L}^2}
\end{equation}
is introduced, with $L=1.0 \cdot 10^2$ the dimensionless edge length of the square computational domain and $N$ the number of agents in the computational domain, ranging from $N=4.0 \cdot 10^3$ for $\phi = 0.1$ to $N=3.2\cdot10^4$ for $\phi=0.8$.

%%%%%%%%%%%%%%%
\begin{figure*}[t!]
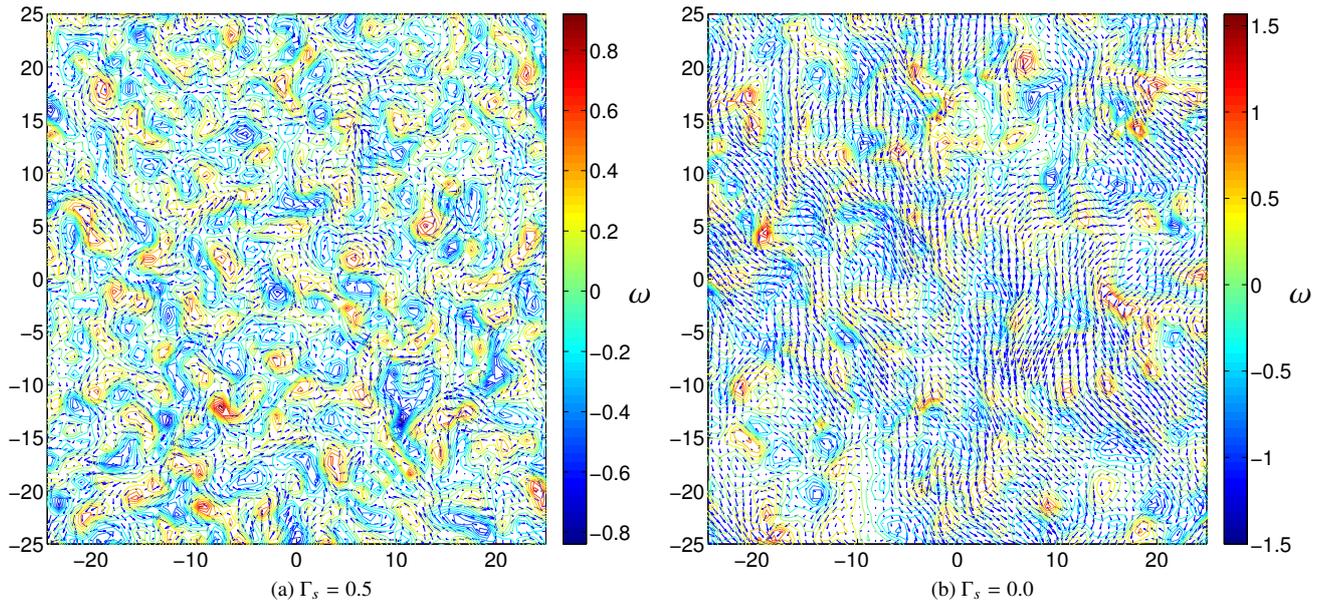

\begin{center}
%\begin{picture}(500,200)

\graphicspath{ {./graphics/serie07_vel_postprocessing/postResults/serie07_T_0p05_prop_1p0_r_0p5_a1p5_g0p5//} }
\subfloat[$\Gamma_s = 0.5$\label{subfig-1:dummy}]{%
\includegraphics[width=.45\textwidth] {phi_0p8_l_0p75_Gamma_8p0/VEL_small2000000_128.pdf}}
\graphicspath{ {./graphics/serie07_vel_postprocessing/postResults/serie07_T_0p05_prop_1p0_r_0p5_a1p5_g0p0//} }
\put(-5, 100){ \large $\omega$ }
$\quad$
\subfloat[$\Gamma_s = 0.0$\label{subfig-1:dummy}]{%
\includegraphics[width=.45\textwidth] {phi_0p8_l_0p75_Gamma_8p0/VEL_small2000000_128.pdf}}
\put(-5, 100){ \large $\omega$ }
\end{center}
\caption{Snapshots of the velocity vectors and vorticity contours of the projected and filtered velocity field $\bfu_M$ in a $50\times50$ region in the center of the computational domain for $\phi = 0.8$ and $\Gamma = 8.0$. The velocity vectors are scaled such that the largest magnitude of the velocity field corresponds to a vector length of 2.0 dimensionless length units. The color coding is chosen such that dark blue corresponds to the smallest negative value in the domain and red corresponds to the largest positive value in the domain. Notice the different color coding in panels (a) and (b). Panel (a): $\Gamma_s = 0.5$. Panel (b): $\Gamma_s = 0.0$.}
\label{fig:velField01}
\end{figure*}
%%%%%%%%%%%%%%%

%%%%%%%%%%%%%%%%%%%%%%%%%%%%%%%%%%%%%%%%%%%%%%%%
\subsection{Agent distributions and velocity fields}
%%%%%%%%%%%%%%%%%%%%%%%%%%%%%%%%%%%%%%%%%%%%%%%%

The influence of the pairwise dissipative interactions along with the volume fraction on the agent distributions is now considered.  
Figure~\ref{fig:particleField01} displays the agent distribution fields with various values of $\phi$ for a fixed value of $\Gamma=8.0$ and four choices of the Stokes friction parameter, namely $\Gamma_s=0.5$, $\Gamma_s=0.3$, $\Gamma_s=0.1$, and the limiting case of $\Gamma_s=0.0$. Visual inspection of the agent distribution fields indicates that, with decreasing Stokes friction and in the limiting case of vanishing Stokes friction $\Gamma_s=0.0$ (dominating pairwise dissipative interactions), the system develops agent density fluctuations with regions of high-density agent aggregation and regions with lower agent concentration. As might be expected, this effect is more prominent with increasing area fraction $\phi$. 

A qualitative understanding of the spatial structures of the velocity fields of the self-propelled agents is provided by considering a filtered and projected velocity field $\bfu_M$. The filtering and projection procedure is explained in detail in Section~\ref{sec:filtESP}. Vortical structures in the velocity field $\bfu_M$ are studied by considering the scalar measure   
\begin{equation}
\omega = \bfomega \cdot \bfe_z
\end{equation}
of vorticity, where $\bfe_z$ is a unit vector normal to the computational domain and $\bfomega$ the vorticity 
\begin{equation}
\bfomega = \Curl \bfu_M.
\end{equation}
An illustrative qualitative comparison of the velocity and vorticity fields between the case $\Gamma_s=0.5$ and the limiting case $\Gamma_s=0.0$ is provided in Figure~\ref{fig:velField01}, showing snapshots of the filtered and projected velocity field $\bfu_M$ along with contours of the scalar vorticity $\omega$ in a $50\times 50$ region of the computational domain. Visual inspection of Figure~\ref{fig:velField01} suggests that velocity fields of both cases exhibit vortical structures of different characteristic size, specific to each case. The limiting case of $\Gamma_s=0.0$ leads to larger structures with correlated velocities over spatially extended regions of greater characteristic size, as can be seen from the scalar vorticity contours and the velocity vectors shown in Figure~\ref{fig:velField01}~(b). Further, the limiting case of $\Gamma_s=0.0$ exhibits larger positive and negative peak magnitudes of $\omega$, as is evident from the different color scales in Figure~\ref{fig:velField01}~(a) and (b).

%%%%%%%%%%%%%%%
\begin{figure*}[!t]
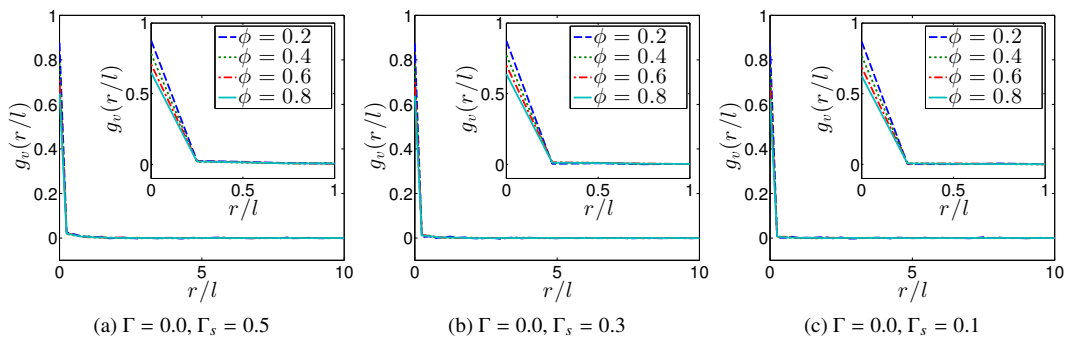

\begin{center}
\graphicspath{ {./graphics/serie07_plots/graphics_1800000_2000000_20000/serie07_T_0p05_prop_1p0_r_0p5_a1p5_g0p5//} }
\subfloat[$\Gamma=0.0$, $\Gamma_s=0.5$\label{subfig-1:dummy}]{%
\includegraphics[width=.25\textwidth] {VCFN_gamma_0p0inset}}
\graphicspath{ {./graphics/serie07_plots/graphics_1800000_2000000_20000/serie07_T_0p05_prop_1p0_r_0p5_a1p5_g0p3//} }
\subfloat[$\Gamma=0.0$, $\Gamma_s=0.3$\label{subfig-2:dummy}]{%
\includegraphics[width=.25\textwidth] {VCFN_gamma_0p0inset}}
\graphicspath{ {./graphics/serie07_plots/graphics_1800000_2000000_20000/serie07_T_0p05_prop_1p0_r_0p5_a1p5_g0p1//} }
\subfloat[$\Gamma=0.0$, $\Gamma_s=0.1$\label{subfig-2:dummy}]{%
\includegraphics[width=.25\textwidth] {VCFN_gamma_0p0inset}}
%
%\put(-10,10){ $\Gamma_s=1.0$ }
\end{center}
\caption{Spatial velocity correlation function for systems with Stokes friction only for different values of the Stokes friction parameter: (a) $\Gamma_s= 0.5$, (b) $\Gamma_s= 0.3$, and (c) $\Gamma_s = 0.1$.  }
\label{fig:VCF}
\end{figure*}
%%%%%%%%%%%%%%%

%%%%%%%%%%%%%%%
\begin{figure*}[!t]
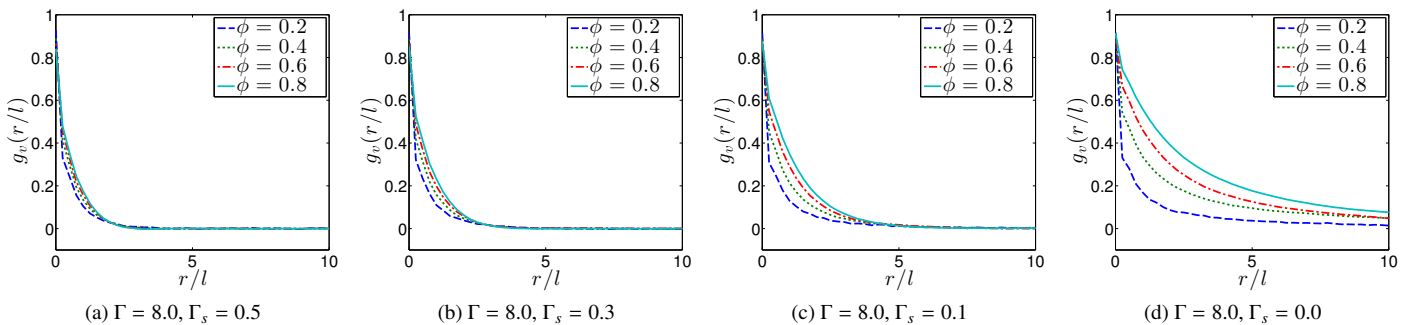

\graphicspath{ {./graphics/serie07_plots/graphics_1800000_2000000_20000/serie07_T_0p05_prop_1p0_r_0p5_a1p5_g0p5//} }
\subfloat[$\Gamma=8.0$, $\Gamma_s=0.5$\label{subfig-1:dummy}]{%
\includegraphics[width=.24\textwidth] {VCFN_gamma_8p0}}
\hfill
\graphicspath{ {./graphics/serie07_plots/graphics_1800000_2000000_20000/serie07_T_0p05_prop_1p0_r_0p5_a1p5_g0p3//} }
\subfloat[$\Gamma=8.0$, $\Gamma_s=0.3$\label{subfig-2:dummy}]{%
\includegraphics[width=.24\textwidth] {VCFN_gamma_8p0}}
\hfill
\graphicspath{ {./graphics/serie07_plots/graphics_1800000_2000000_20000/serie07_T_0p05_prop_1p0_r_0p5_a1p5_g0p1//} }
\subfloat[$\Gamma=8.0$, $\Gamma_s=0.1$\label{subfig-2:dummy}]{%
\includegraphics[width=.24\textwidth] {VCFN_gamma_8p0}}
\hfill
\graphicspath{ {./graphics/serie07_plots/graphics_1800000_2000000_20000/serie07_T_0p05_prop_1p0_r_0p5_a1p5_g0p0//} }
\subfloat[$\Gamma=8.0$, $\Gamma_s=0.0$\label{subfig-2:dummy}]{%
\includegraphics[width=.24\textwidth] {VCFN_gamma_8p0}}
\caption{Spatial velocity correlation function for systems with Stokes friction and pairwise dissipative interactions (panels (a)--(c)), and for systems with pairwise dissipative interactions only (panel (d)).}
\label{fig:VCF02}
\end{figure*}
%%%%%%%%%%%%%%%

%%%%%%%%%%%%%%%%%%%%%%%%%%%%%%%%%%%%%%
\subsection{Two-point velocity correlation function}
%%%%%%%%%%%%%%%%%%%%%%%%%%%%%%%%%%%%%%

A more quantitative understanding of the spatial structure of the velocity field is provided by the normalized one-time spatial two-point velocity correlation function (VCF) $g_v$ defined by (see, for example, Cisneros et al.~\cite{Cisneros2007a})
\begin{equation}\label{eq:VCF}
g_v(r) = \frac{ \langle \bfv_{\rm{COM}}(\bfx) \cdot \bfv_{\rm{COM}}(\bfx+\bfr) \rangle_r - \langle \bfv_{\rm{COM}} \rangle_r^2 }{ \langle \bfv_{\rm{COM}}^2 \rangle_r - \langle \bfv_{\rm{COM}} \rangle_r^2},
\end{equation}
with $\bfv^{k}_{\rm{COM}} = \half (\bfv^{(k_1)} +\bfv^{(k_2)})$ the dimensionless center of mass (COM) velocity of agent $k$ and $\langle \cdot \rangle_r$ the spatial average over the computational domain conditioned on the separation distance $r$. Specifically, given a quantity $a$ depending on $\bfx$ and $\bfr$, 
\begin{equation}\label{eq:spatial01}
\langle a(\bfx,\bfr) \rangle_r = \frac{1}{N_r} \sum_{|\bfr|=r} a(\bfx,\bfr) 
\end{equation}
denotes its spatial average over all agents with separation distance $r$ in the computational domain, where $N_r$ is the number of agents satisfying the condition $|\bfr|=r$. The VCF is a measure of the correlation of the velocity field over different length scales given through the separation distance $r$. For example, the limit $g_v(r=R) \rightarrow 1.0$ corresponds to a velocity field with statistically perfectly correlated velocities on length scales of linear dimension $R$. 
On the other hand, the limit $g_v(r=R) \rightarrow 0.0$ corresponds to a velocity field with statistically uncorrelated velocities on length scales of linear dimension $R$. Heuristically, if a velocity field exhibits a high correlation at a certain length scale $R$, this means that on average spatial regions of size $R$ move with the same velocity, forming spatial structures of size $R$. 

The VCFs of three different types of systems, representing the previously discussed limiting cases for the two friction parameters $\Gamma$ and $\Gamma_s$ for selected representative volume fractions are considered. In particular, the VCFs of systems with Stokes friction only are shown in Figure~\ref{fig:VCF}, the VCFs of three systems with Stokes friction and pairwise dissipative interactions are shown in panels (a)--(c) of Figure~\ref{fig:VCF02}, and the VCFs of a system with pairwise dissipative interactions only are shown in panel (d) of Figure~\ref{fig:VCF02}. The weakest correlations, evident through non-zero values of the VCF only for very small separation distances, are found in systems with Stokes friction only, as shown in Figure~\ref{fig:VCF}. Further, systems with Stokes friction and pairwise dissipative interactions exhibit correlated velocities over a broader range of separation distances as shown for three examples of $\Gamma_s = 0.5$, $\Gamma_s = 0.3$, and $\Gamma_s=0.1$ in panels (a)--(c) of Figure~\ref{fig:VCF02}, respectively.  Confirming the observations from visual inspection of the simulation snapshots, in the limiting case of pairwise dissipative interactions only, spatially extended strong correlations are evident up to large separation distances, as shown in panel (d) of Figure~\ref{fig:VCF02}. 
For the case of pairwise dissipative interactions only, the correlations increase consistently with increasing $\phi$ (panel (d) of Figure~\ref{fig:VCF02}), in contrast to the cases of combined Stokes and pairwise dissipative interactions (panels (a)--(c) of Figure~\ref{fig:VCF02}) and Stokes friction only (Figure~\ref{fig:VCF}), where the area fraction has only a weak influence on the VCFs. 

%%%%%%%%%%%%%%%
\begin{figure*}[!t]
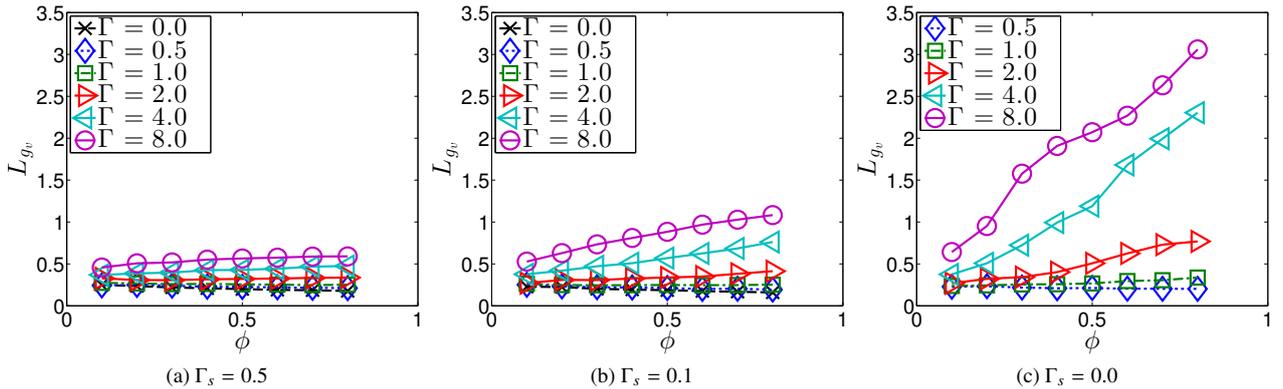

\begin{center}
\graphicspath{ {./graphics/serie07_plots/graphics_1800000_2000000_20000/serie07_T_0p05_prop_1p0_r_0p5_a1p5_g0p5//} }
\subfloat[$\Gamma_s=0.5$\label{subfig-1:dummy}]{%
\includegraphics[width=.3\textwidth] {VCFN_Lgv}}
\graphicspath{ {./graphics/serie07_plots/graphics_1800000_2000000_20000/serie07_T_0p05_prop_1p0_r_0p5_a1p5_g0p1//} }
\subfloat[$\Gamma_s=0.1$\label{subfig-2:dummy}]{%
\includegraphics[width=.3\textwidth] {VCFN_Lgv}}
\graphicspath{ {./graphics/serie07_plots/graphics_1800000_2000000_20000/serie07_T_0p05_prop_1p0_r_0p5_a1p5_g0p0//} }
\subfloat[$\Gamma_s=0.0$\label{subfig-2:dummy}]{%
\includegraphics[width=.3\textwidth] {VCFN_Lgv}}
%
%\put(-10,10){ $\Gamma_s=1.0$ }
\end{center}
\caption{Characteristic integral length scale for systems with Stokes friction and pairwise dissipative interactions (panels (a) and (b)) and systems with pairwise dissipative interactions only (panel (c)) for different values of the area fraction $\phi$. Notice that (a) and (b) includes the limiting cases of Stokes friction only, namely $\Gamma = 0.0$.}
\label{fig:integralScale}
\end{figure*}
%%%%%%%%%%%%%%%

Based on the VCF~\eqref{eq:VCF}, it is useful to define a dimensionless integral length scale (see, for example, Pope~\cite{Pope2000})
\begin{equation}
L_{g_v} = \int_0^\infty g_v(r) \, {\rm{d}}r, 
\end{equation}
which is the (nondimensional) area under the VCF and quantifies the average characteristic linear dimension of spatial structures of the COM velocity field. 
The features of $L_{g_v}$ for the entire range of volume fractions and pairwise friction parameters considered are presented for two cases of Stokes friction and pairwise dissipative interactions ($\Gamma_s =0.5$ and $\Gamma_s =0.1$) and pairwise dissipative interactions only ($\Gamma_s =0.0$), as shown in panels (a), (b), and (c) of Figure~\ref{fig:integralScale}, respectively. Notice that the cases $\Gamma_s=0.5$ and $\Gamma_s=0.1$ shown in panels (a) and (b) of Figure~\ref{fig:integralScale} both include the limiting case of Stokes friction only --- that is, the case $\Gamma = 0.0$. For $\Gamma_s=0.5$, $L_{g_v}$ is nearly constant for different values of the area fractions $\phi$ and increases only weakly with an increase in $\Gamma$, as shown in panel (a) of Figure~\ref{fig:integralScale}. For lower values of Stokes friction, $L_{g_v}$ increases with increasing $\phi$, as shown in panel (b) of Figure~\ref{fig:integralScale}. Consistent with the previously discussed VCFs, the smallest values of $L_{g_v}$ over the entire range of $\phi$ arise if only Stokes friction is active, as shown in the limiting cases of $\Gamma =0.0$ in panels (a) and (b) of Figure~\ref{fig:integralScale}.

In contrast, for $\Gamma_s=0.0$, $L_{g_v}$ significantly increases with increasing $\Gamma$, as shown in panel (c) of Figure~\ref{fig:integralScale}. This effect becomes substantially more prominent with increasing $\phi$. In conclusion, vanishing Stokes friction in combination with large pairwise dissipative interactions lead to spatially extended correlated motion patterns. While present at all considered values of $\phi$, this effect becomes substantially more prominent with increasing $\phi$. Bearing in mind that self-propulsion forces inject energy into the system on a length scale on the order of the dimensionless length  $l=1.0$ of an individual agent, this behaviour points to the presence of upscale energy transfer and energy condensation leading to the formation of spatially extended structures in the COM velocity field.

An explanation for the observed behavior can be found by considering the microscopic interaction mechanism between particle pairs. Dominating pairwise dissipative interactions reduce the elasticity of the forces between particle pairs. For example, consider two approaching particles. Without pairwise dissipative interactions, the particles experience an elastic repulsive force that eventually reverses their direction of motion. However, this behaviour is not observed in actual systems of interacting self-motile bacteria. Instead, short-ranged hydrodynamic interactions between the agents make the interactions inelastic. The model accounts for this mechanism through the simple pairwise dissipative interactions. 

Although pairwise dissipative interactions provide only a very simple model for hydrodynamic interaction forces, the observed behavior suggests that such forces form spatially extended correlated motion patterns in self-propelled particle systems.

%%%%%%%%%%%%%%%
\begin{figure*}[!t]
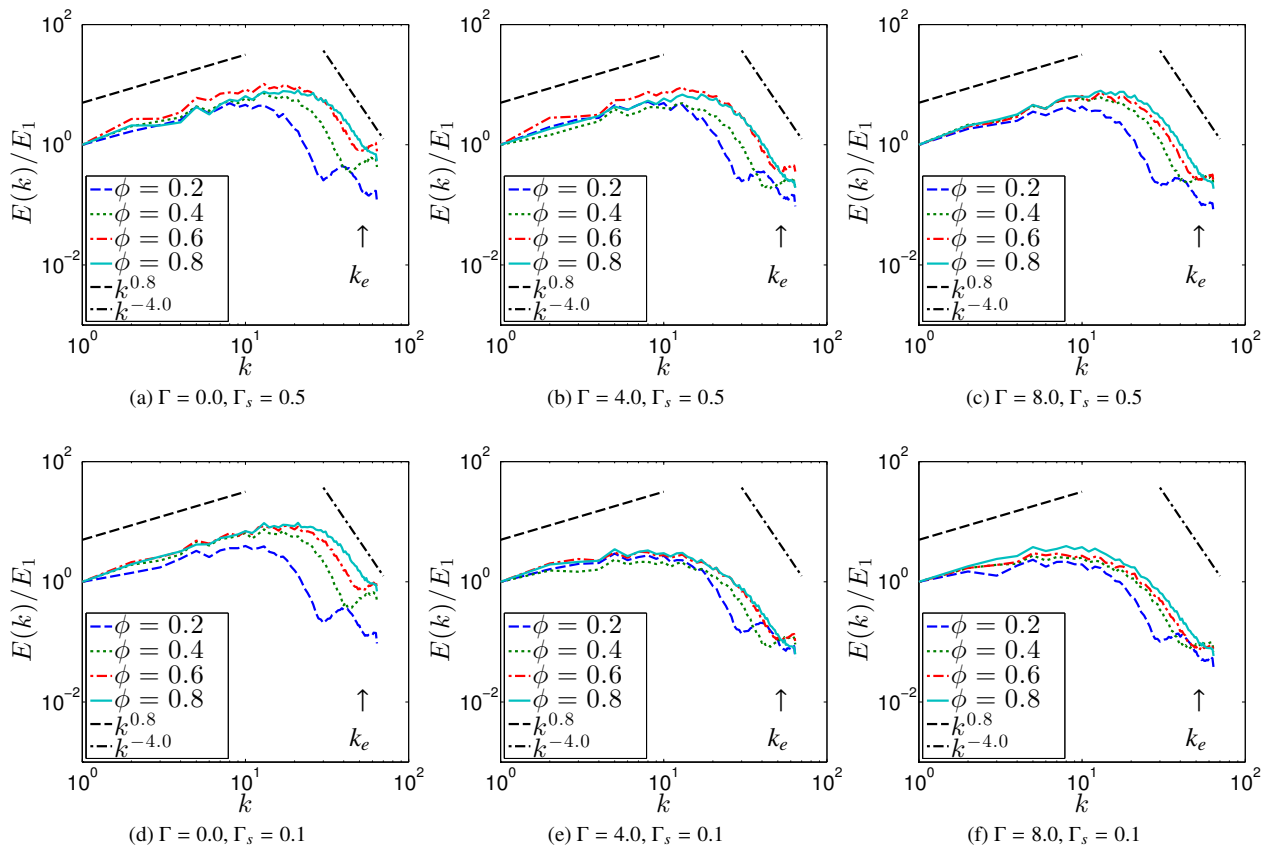

\begin{center}
\graphicspath{ {./graphics/serie07_plots/graphics_1800000_2000000_20000/serie07_T_0p05_prop_1p0_r_0p5_a1p5_g0p5//} }
\subfloat[$\Gamma=0.0$, $\Gamma_s=0.5$\label{subfig-1:dummy}]{%
\includegraphics[width=.3\textwidth] {ESPN03_128_Gamma_0p0}}
\put(-29,50){ $\uparrow$ }
\put(-29,35){$k_e$ }
\subfloat[$\Gamma=4.0$, $\Gamma_s=0.5$\label{subfig-2:dummy}]{%
\includegraphics[width=.3\textwidth] {ESPN03_128_Gamma_4p0}}
\put(-29,50){ $\uparrow$ }
\put(-29,35){$k_e$ }
\subfloat[$\Gamma=8.0$, $\Gamma_s=0.5$\label{subfig-2:dummy}]{%
\includegraphics[width=.3\textwidth] {ESPN03_128_Gamma_8p0}}
\put(-29,50){ $\uparrow$ }
\put(-29,35){$k_e$ }
\graphicspath{ {./graphics/serie07_plots/graphics_1800000_2000000_20000/serie07_T_0p05_prop_1p0_r_0p5_a1p5_g0p1//} }
\subfloat[$\Gamma=0.0$, $\Gamma_s=0.1$\label{subfig-1:dummy}]{%
\includegraphics[width=.3\textwidth] {ESPN03_128_Gamma_0p0}}
\put(-29,40){ $\uparrow$ }
\put(-29,25){$k_e$ }
\subfloat[$\Gamma=4.0$, $\Gamma_s=0.1$\label{subfig-2:dummy}]{%
\includegraphics[width=.3\textwidth] {ESPN03_128_Gamma_4p0}}
\put(-29,40){ $\uparrow$ }
\put(-29,25){$k_e$ }
\subfloat[$\Gamma=8.0$, $\Gamma_s=0.1$\label{subfig-2:dummy}]{%
\includegraphics[width=.3\textwidth] {ESPN03_128_Gamma_8p0}}
\put(-29,40){ $\uparrow$ }
\put(-29,25){$k_e$ }
\end{center}
\caption{Normalized two-dimensional energy spectra for systems with Stokes friction and pairwise dissipative interactions, including the limiting case of Stokes friction only in panels (a) and (d). Panels (a)--(c) are for $\Gamma_s = 0.5$ and panels (d)--(f) are for $\Gamma_s=0.1$. }
\label{fig:ESPN01}
\end{figure*}
%%%%%%%%%%%%%%%

%%%%%%%%%%%%%%%%%%%%%%%%%%%%%%%%%%%%%%
\subsection{Filtered and projected velocity fields and energy spectra}
\label{sec:filtESP}
%%%%%%%%%%%%%%%%%%%%%%%%%%%%%%%%%%%%%%

To explore the energetic properties of the system, two-dimensional kinetic-energy spectra are considered. To compute these spectra, the particle velocity fields $\bfv^{i}$ at all particle locations $\bfx^{i}$ are filtered and projected onto a uniform grid with equidistant spacing in both coordinate directions. The uniform grid is comprised by a set of discrete spatial points $\bfx_M$, where $M$ is the number of uniformly spaced grid points in each coordinate direction. Further, let $\bfu_M(\bfx_M)$ denote the filtered and projected velocity field at a spatial point $\bfx_M$ on the uniform grid, as computed in terms of the particle velocity fields  $\bfv^{i}$ through the convolution filter operation
\begin{equation}
\bfu_M(\bfx_M) = \frac{1}{n} \sum\limits_{i=1}^N \bfv^{i}\psi(\bfx_M - \bfx^{i}),
\end{equation}
where $\psi$ is a sufficiently rapidly decaying filtering kernel and $n$ the number of particles for which $\psi \neq 0$. 
In the present study, a square grid filter
\begin{equation} 
\psi(\bfr) = \left\{\begin{array}{lll}  1.0,   & \bfr \cdot \bfe_x \le \epsilon \quad \text{and}  & \bfr \cdot \bfe_y \le \epsilon, \\  
%									 0,  & \bfr \cdot \bfe_x > \epsilon \quad \text{or} & \bfr \cdot \bfe_y > \epsilon, \end{array} \right .
									 0.0,  & \text{otherwise},  \end{array} \right .
\end{equation}
with $\epsilon$ the nondimensional filter width, is used to filter and project the particle velocity field. Whereas the special case of $\epsilon = \Delta x$, with $\Delta x = L/M$ the uniform spacing between grid points, represents a non-overlapping grid filter, a choice of $\epsilon > \Delta x$ results in overlapping filter bins and, thus, is referred to as an overlapping grid filter. In the present study, an overlapping grid filter is considered and the bin size is determined based on the desired average number of agents $N_f$ in each filter bin. Importantly, $\epsilon$, $L$, and $N$ are related by 
\begin{equation}\label{eq:filterwidth}
\epsilon = L \sqrt{\frac{N_f}{N}},
\end{equation}
where $N_f$ is the desired average number of agents in each filter bin. 
To compute meaningful averages, $\epsilon$ is chosen according to~\eqref{eq:filterwidth} and such that that every filter bin contains approximately $N_f=10$ agents. 

To estimate the number of points of the uniform grid necessary to resolve the energy spectrum in sufficient detail, consider the approximate length scale $l_e$ of small-scale energy injection through the self-propulsion forces. This scale may be estimated to be on the order of the length of an individual bacterium --- that is, $l_e \sim O(l)$. To capture the motion at length scales on the order of $l_e$, it is thus required that the grid spacing $\Delta x$  used to compute the energy spectrum obeys $\Delta x < l_e$. Bearing in mind the ratio $L/l = 1.0\cdot 10^2$, the number of points $M$ in each coordinate direction on the uniform grid must therefore obey $M>100$. For the present purpose, it is assumed that $M=128$.

Let $\hat{\bfu}_M$ denote the complex Fourier coefficient corresponding to $\bfu_M$, so that 
\begin{equation}\label{eq:fourierTrans01}
\bfu_M(\bfx_M) = \sum_{\cal{I}} \hat{\bfu}_M(\bfk) {\text{e}}^{i \bfk \cdot \bfx_M},
\end{equation}
with $\cal{I}$ a finite collection of Fourier modes and $\bfk$ the two-dimensional normalized wavenumber vector. In view of~\eqref{eq:fourierTrans01}, the two-dimensional kinetic-energy spectrum $E$ of the filtered velocity field $\bfu_M$ is defined as
\begin{equation}\label{eq:ESPf}
E(k) = \int_{|\bfk|=k} \tfrac{1}{2}|\hat{\bfu}_M(\bfk)|^2 \,\text{d}\bfk,
\end{equation}
where $k=|\bfk|$ is the magnitude of $\bfk$. The choice $M^2 = 128^2$ of the number of points on the uniform grid corresponds to the maximum representable normalized wavenumber $k_{\text{max}}=64$ and the energy injection lengthscale $l_e=l$ corresponds to the normalized energy injection wavenumber $k_e = 50$. 

To investigate the relevance of power-law scalings, the energy spectrum is normalized by the energy contained in the first wavenumber shell $E_1$. 
For two- and three-dimensional flows of Newtonian fluids, energy spectra exhibit power-law scalings of the general form
\begin{equation}\label{eq:powerLaw01}
E(k)/E_1 \propto C k^{\alpha},
\end{equation}
with $C$ and $\alpha$ dimensionless constants. In particular, $\alpha=-5/3$ corresponds to Kolmogorov's scaling for the inertial range of three-dimensional homogeneous and isotropic Navier--Stokes (NS) turbulence (see, for example, Frisch~\cite{Frisch1996}). Notice that the VCF and the energy spectrum form a Fourier transform pair and that the intermediate wavenumber power-law slope of the energy spectrum $E/E_1$ is related to the integral length scale of the VCF (see, for example, Pope~\cite{Pope2000}). 

%%%%%%%%%%%%%%%
\begin{figure*}[!t]
\begin{center}
\graphicspath{ {./graphics/serie07_plots/graphics_1800000_2000000_20000/serie07_T_0p05_prop_1p0_r_0p5_a1p5_g0p0//} }
\subfloat[$\Gamma=1.0$, $\Gamma_s=0.0$\label{subfig-1:dummy}]{%
\includegraphics[width=.3\textwidth] {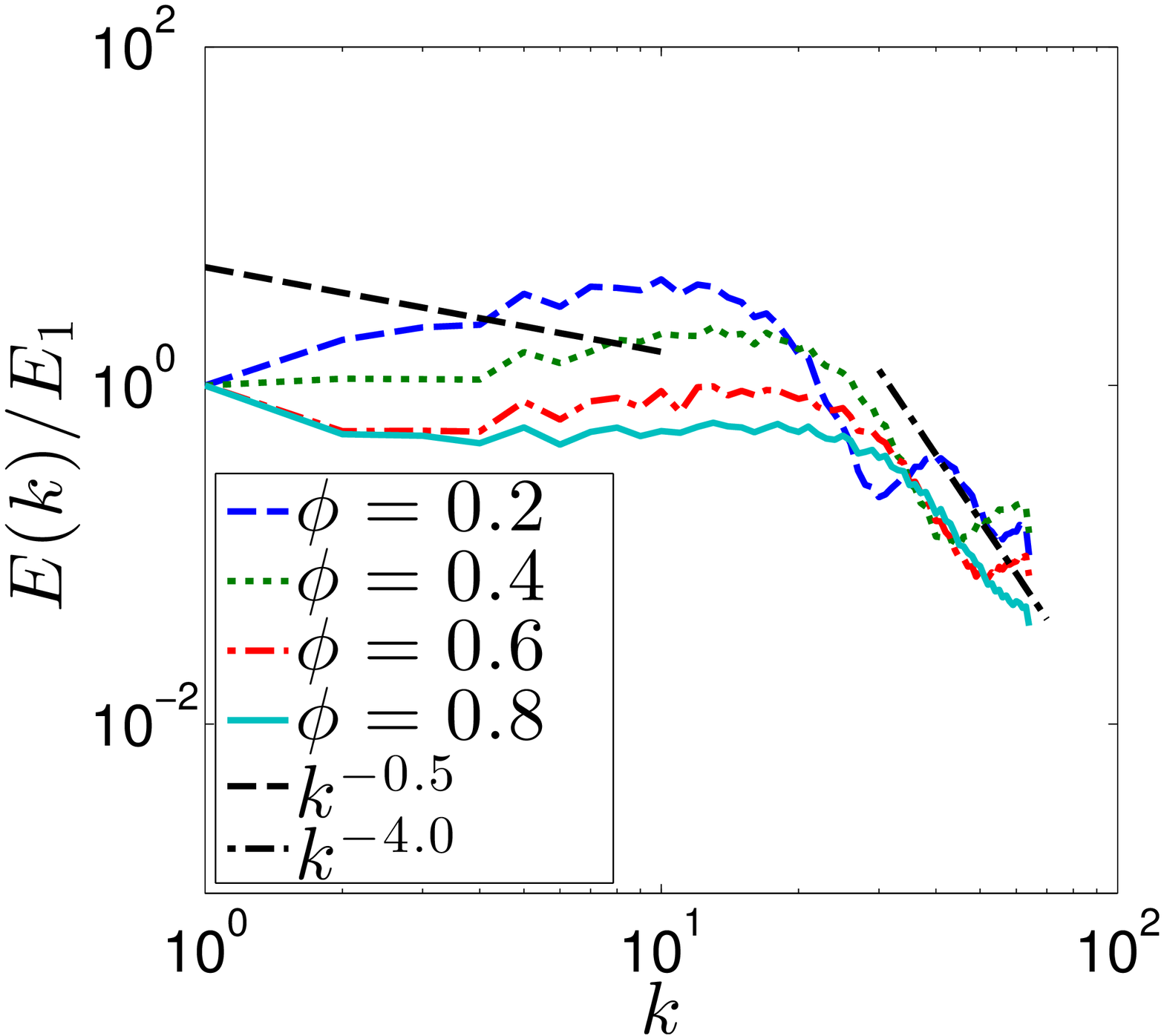}}
\put(-29,25){ $\uparrow$ }
\put(-29,8){$k_e$ }
\subfloat[$\Gamma=4.0$, $\Gamma_s=0.0$\label{subfig-2:dummy}]{%
\includegraphics[width=.3\textwidth] {ESPN03_128_Gamma_4p0}}
\put(-29,25){ $\uparrow$ }
\put(-29,8){$k_e$ }
\subfloat[$\Gamma=8.0$, $\Gamma_s=0.0$\label{subfig-2:dummy}]{%
\includegraphics[width=.3\textwidth] {ESPN03_128_Gamma_8p0}}
\put(-29,25){ $\uparrow$ }
\put(-29,8){$k_e$ }
\end{center}
\caption{Normalized two-dimensional energy spectra for systems with pairwise dissipative interactions only ($\Gamma_s=0.0$) for different values of the pairwise dissipative interaction force parameter: (a) $\Gamma= 1.0$, (b) $\Gamma= 4.0$, and (c) $\Gamma = 8.0$. }
\label{fig:ESPN02}
\end{figure*}
%%%%%%%%%%%%%%%

In classical turbulence the discussion of spectral energy distribution and the corresponding scaling laws is usually accompanied by a notion of scale separation (see, for example, Pope~\cite{Pope2000}). In the spectral distribution of energy low wavenumber contributions are associated with an integral range, intermediate wavenumber contributions are associated with an inertial range, and small wavenumber contributions are associated with a dissipation range. Despite a tentative resemblance between the statistics for the energy spectrum $E/E_1$ obtained in the present setting and phenomena observed in classical turbulence, the notion of scale separation in the context of active suspensions is unorthodox. It therefore seems reasonable to distinguish between large, intermediate, and small scales associated with low, intermediate, and high wavenumbers, respectively, rather than integral, inertial, and dissipation ranges (as is conventional in discussion of turbulent fluid flow). 

In the present investigation, power-law scaling exponents are computed with a least-square fit of the energy spectrum $E/E_1$  to the power law~\eqref{eq:powerLaw01} in the interval of the first 10 wavenumber shells (that is, $k \leq 10$).

\begin{figure*}[!t]
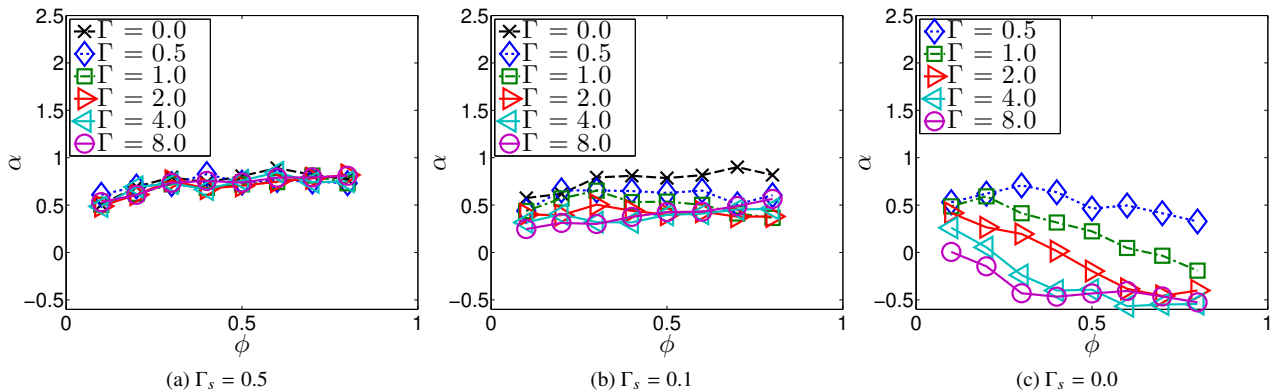

\begin{center}
\graphicspath{ {./graphics/serie07_plots/graphics_1800000_2000000_20000/serie07_T_0p05_prop_1p0_r_0p5_a1p5_g0p5//} }
\subfloat[$\Gamma_s=0.5$\label{subfig-1:dummy}]{%
\includegraphics[width=.3\textwidth] {ESP03_N128_slope}}
\graphicspath{ {./graphics/serie07_plots/graphics_1800000_2000000_20000/serie07_T_0p05_prop_1p0_r_0p5_a1p5_g0p1//} }
\subfloat[$\Gamma_s=0.1$\label{subfig-2:dummy}]{%
\includegraphics[width=.3\textwidth] {ESP03_N128_slope}}
\graphicspath{ {./graphics/serie07_plots/graphics_1800000_2000000_20000/serie07_T_0p05_prop_1p0_r_0p5_a1p5_g0p0//} }
\subfloat[$\Gamma_s=0.0$\label{subfig-2:dummy}]{%
\includegraphics[width=.3\textwidth] {ESP03_N128_slope}}
\end{center}
\caption{Power-law scaling slope $\alpha$ at low wavenumber of the normalized energy spectrum for systems with Stokes friction and pairwise dissipative interactions  (panels (a) and (b)) and systems with pairwise dissipative interactions  only (panel (c)) for different values of the pairwise friction coefficient $\Gamma$ and different area fractions $\phi$. Notice that panels (a) and (b) include the limiting case of Stokes friction only, that is the case $\Gamma_s=0.0$ of vanishing pairwise friction. The slope was computed with a least-squares fit to the interval of the normalized energy spectrum of the smallest ten wavenumber shells. }
\label{fig:ESPScale}
\end{figure*}

The influences of Stokes friction only and mixed Stokes and pairwise dissipative interactions on $E/E_1$ are presented in Figure~\ref{fig:ESPN01}. In particular, plots of $E/E_1$ are provided for different combinations of Stokes friction and pairwise dissipative interactions. Two different cases of Stokes friction are displayed, namely $\Gamma_s =0.5$ in panels (a)--(c) and $\Gamma_s =0.1$ in panels (d)--(f). The pairwise dissipative interaction force includes the limiting case of $\Gamma=0.0$ and two mixed cases ($\Gamma_s=0.5$ and $\Gamma=4.0$, $8.0$). While the energy spectra exhibit a positive scaling law with the scaling exponent $\alpha \approx 0.8$ at low wavenumbers, they quickly decrease at high wavenumbers with an approximate power-law slope of $\alpha_D \approx -4.0$. These results indicate that the low and high wavenumber behviors appear to be universal in the sense that all the considered cases feature a low wavenumber power-law scaling with $E/E_1 \propto k^{0.8}$ and a high wavenumber scaling with $E/E_1 \propto k^{-4.0}$. This result is confirmed in more detail in panel (a) and (b) of Figure~\ref{fig:ESPScale}. The figure displays the low wavenumber scaling exponents $\alpha$ versus the area fraction $\phi$ for $\Gamma_s=1.0$ with various values of $\Gamma$ ranging from $0.0$ to  $8.0$. Strikingly, the low wavenumber scaling exponent $\alpha$ is approximately constant for all considered pairwise dissipative interaction parameters $\Gamma$, including the limiting case $\Gamma = 0.0$, with a weak decrease for very small values of the area fraction $\phi$. 

Furthermore, the positive low wavenumber scaling of the cases with Stokes friction is accompanied by a global maximum of the energy spectrum at intermediate wavenumbers around $k\approx 20$. The exact location of this maximum appears to depend on the volume fraction of the system, with higher volume fractions leading to peaks at larger wavenumbers. Since the energy injection through self-propulsion is associated with a band of high wavenumbers around $k_e\approx 50$ (marked with an arrow in Figure~\ref{fig:ESPN01}), the peak at $k\approx 20 $ corresponds to an accumulation of energy at the corresponding intermediate length scales. In the case of large Stokes friction ($\Gamma_s=0.5$) and pairwise dissipative interactions, this accumulation of energy appears to be essentially independent of the choice of $\Gamma$. For lower Stokes friction ($\Gamma_s=0.1$), pairwise dissipative interactions have more influence on the energy spectra, moving the peak in the energy spectrum to lower wavenumbers, as shown in panels (d)--(e) of Figure~\ref{fig:ESPN01}.

%%%%%%%%%%%%%%%
%%%%%%%%%%%%%%%

Figure~\ref{fig:ESPN02} shows energy spectra with the objective of investigating the influence of pairwise friction parameter $\Gamma$ by setting $\Gamma_s=0.0$. Plots are displayed for the cases of $\Gamma=1.0$, $4.0$, $8.0$ and $\phi=0.2$, $0.4$, $0.6$, $0.8$. For all three choices of $\Gamma$, the energy spectra exhibit low wavenumber scalings that are significantly more sensitive to the area fraction than in the cases with Stokes friction. In contrast, the high wavenumber range exhibits a  scaling law similar to that arising in the cases with Stokes friction, with $\alpha_D \approx -4.0$. As $\phi$ and $\Gamma$ increase, the slope of the energy spectrum at low wavenumbers decreases. In particular, for $\Gamma=1.0$, the slope at the low wavenumbers more strongly depends on $\phi$  than larger values of $\Gamma$, as is evident from panel (a) of Figure~\ref{fig:ESPN02}. Interestingly, while all energy spectra for $\Gamma=8.0$ show a low wavenumber scaling with a negative scaling exponent, as shown in panel (c) of Figure~\ref{fig:ESPN02}, the scaling exponent of the energy spectra for the other choices of $\Gamma$ changes from positive to negative with increasing $\phi$.  Panel (c) of Figure~\ref{fig:ESPScale}, provides the scaling exponents obtained from a least-squares fit of the energy spectra. These results further confirm that the pairwise friction parameter $\Gamma$ has a significant impact on the low wavenumber scaling of the normalized energy spectrum; lower values of $\Gamma$ result in positive slopes and higher values of $\Gamma$ combined with high area fractions $\phi$ lead to negative low wavenumber scaling slopes, where $\alpha \approx - 0.5 $ appears to be the lower bound observed for the current combinations of parameters.

In agreement with the previously discussed VCF statistics and the representative snapshots of the velocity and vorticity fields, the case of zero Stokes friction induces a distinct behaviour of the normalized energy spectrum $E/E_1$. For low values of the pairwise friction parameter $\Gamma$, the low wavenumber slope of $E/E_1$ strongly depends on the area fraction $\phi$, as shown in panel (a) of Figure~\ref{fig:ESPN02}. Increasing $\phi$ leads to a decrease in the slope of $E/E_1$. Further, increasing $\Gamma$ leads to a decrease in the slope at low wavenumbers of $E/E_1$ for all considered values of $\phi$, as shown in panels (a)--(c) of Figure~\ref{fig:ESPScale}. The observed decrease in the slope of the low and interemediate wavenumber range of $E/E_1$ is more prominent for large $\phi$; however, for the largest considered pairwise friction coefficient, namely $\Gamma = 8.0$, all normalized energy spectra $E/E_1$ possess a low wavenumber scaling with a negative scaling exponent $\alpha$, as shown in panel (b) of Figure~\ref{fig:ESPScale}.

The behaviour of the normalized energy spectra $E/E_1$ point to the presence of different energy transfer mechanisms in the cases with and without Stokes friction. As is apparent from Figure~\ref{fig:ESPN01}, the energy spectra with Stokes friction exhibit a global maximum at intermediate wavenumbers for all considered values of the area fraction and pairwise friction parameters.  However, without Stokes friction, the energy spectra exhibit a global maximum at intermediate wavenumbers only if both $\Gamma$ and $\phi$ are sufficiently small, as depicted in Figure~\ref{fig:ESPN02}. In addition, for larger values of $\Gamma$ and $\phi$, the normalized energy spectrum of systems without Stokes friction decreases monotonically with increasing wavenumber and attains its global maximum at the smallest wavenumber shell $k=1$. 

These results indicate that, in self-propelled agents without the effect of Stokes friction (that is, $\Gamma_s=0.0$), the energy, after being injected at large wavenumbers cascades upwards to the smallest wavenumber shell, corresponding to the largest length scale in the system. This effect becomes more predominant as the influence of pairwise dissipative interactions and the area fraction increases. Adding the effect of Stokes friction results in the accumulation of the energy at an intermediate wavenumber, forming a peak in the normalized energy spectrum, irrespective of the value of $\Gamma$ or $\phi$. 

The discussed behavior can be understood by considering the spectral distribution of the two different energy dissipation mechanisms. 

\begin{itemize}
\item The energy dissipation due to the Stokes friction is proportional to the (constant) Stokes friction coefficient and the magnitude of the squared velocity, which is twice the (dimensionless) kinetic energy. Consequently, from a spectral perspective, the dissipation due to Stokes friction acts mainly on wavenumbers containing large amounts of kinetic energy, namely the low wavenumbers. In two-dimensional flow, Stokes friction, or similar types of friction mechanisms, for example Rayleigh friction, therefore remove energy on large scales, that is, low wavenumber components of the velocity field (see, for example, Boffetta and Ecke \cite{Boffetta2012}). Such dissipation mechanisms penalize the formation of spatially extended correlated structures in the velocity field.

\item The pairwise dissipative interactions depend on local velocity differences rather than on the magnitude of the velocity. Heuristically, they are thus expected to remove energy at very small scales, allowing for spatially extended correlated motion if the system is not over-damped by dissipation due to Stokes friction. For the limiting case of vanishing Stokes friction, this explains the observed spatially extended correlated motion patterns.

\end{itemize}

%%%%%%%%%%%%%%%%%%%%%%%%%%%%%%%%%%%%%%
\section{Summary and conclusions}
\label{sec:SPPsummary}
%%%%%%%%%%%%%%%%%%%%%%%%%%%%%%%%%%%%%%

A system of self-propelled soft-core dumbbells interacting with DPD-type forces was studied. More particularly, the influences of agent concentration and two different friction mechanisms, namely pairwise dissipative interactions and Stokes friction, on the statistics of the system were presented. The pairwise dissipative forces provide a simple model system for studying the influence of hydrodynamic interactions in active suspensions. 

High agent concentrations combined with dominant pairwise dissipative interactions and vanishing Stokes friction result in dynamic particle aggregation  and spatially extended correlated motions. The characteristic length scales of the spatial structures exceed the characteristic size of an individual agent and point at the presence of phenomena reminiscent of the upscale energy transfer and energy condensation phenomena observed in classical two-dimensional turbulent flows, as discussed, for example, by Boffetta and Ecke~\cite{Boffetta2012}. 

For cases with Stokes friction, the normalized energy spectra possess a small wavenumber scaling with positive scaling exponent and a peak at intermediate wavenumbers. In contrast, the cases with dominant pairwise dissipative interactions have low wavenumber scalings with negative scaling exponents, pointing at the presence of an accumulation of energy at large scales.

Since the formulation includes neither ad-hoc biological alignment and coordination rules nor attractive forces, the obtained results demonstrate the potential importance of pairwise dissipative interactions in the formation of spatially extended structures in active suspensions. Further, the soft-core interaction potentials used in DPD afford numerical stability and efficiency, even at large integration timesteps. Larger timesteps enable efficient and robust simulations, allowing for the consideration of scenarios more complex than would be possible otherwise. 

At the present stage, only one simple form of pairwise dissipative hydrodynamic interactions has been considered. The investigation of different forms of such hydrodynamic interactions should be subject of future research. Possible changes to the hydrodynamic interactions include modifying the weighting function~\eqref{eq:weightFunc01} through changing the exponent with which $w$ enters the DPD interactions~\eqref{eq:DPDcdr}. This possibility is discussed by Pan et al.~\cite{Pan2008}, Fan et al.~\cite{Fan2006}, Fedosov et al.~\cite{Fedosov2008a}, and Symeonidis et al.~\cite{Symeonidis2006}, with the goal of adapting bulk rheological properties like the viscosity or the Schmidt number of a classical DPD fluid. Alternatively, a generalization of $w$ might account for lubrication-type interactions. However, the asymptotic behavior of lubrication interactions as the separation distance vanishes can be expected to compromise part of the stability of the standard soft-core DPD method.

\section*{Acknowledgements}

DFH acknowledges the partial support of the Antje Graupe Pryor Foundation and the Graduate Research Mobility Award of the Department of Mechanical Engineering at McGill University along with the hospitality of the Department of Mathematics at Washington State University.



%\bibliography{/Users/dns/Dropbox/BibTexLibrary/library}
%\bibliography{/Users/denishinz/Dropbox/BibTexLibrary/library}

%\bibliographystyle{elsarticle-num-names}






% \end{thebibliography}

\end{document}